\documentclass[aps,preprint,floats,epsf,epsfig,nofootinbib,letter]{revtex4}

\usepackage{graphicx}
\usepackage{dcolumn}
\usepackage{bm}

\begin{document}
\def\be{\begin{eqnarray}}
\def\en{\end{eqnarray}}
\def\non{\nonumber}
\def\la{\langle}
\def\ra{\rangle}
\def\nc{N_c^{\rm eff}}
\def\vp{\varepsilon}
\def\A{{\cal A}}
\def\B{{\cal B}}
\def\c{{\cal C}}
\def\d{{\cal D}}
\def\e{{\cal E}}
\def\p{{\cal P}}
\def\t{{\cal T}}
\def\N{{\cal N}}
\def\up{\uparrow}
\def\dw{\downarrow}
\def\vma{{_{V-A}}}
\def\vpa{{_{V+A}}}
\def\smp{{_{S-P}}}
\def\spp{{_{S+P}}}
\def\J{{J/\psi}}
\def\ol{\overline}
\def\ov{\overline}
\def\Lqcd{{\Lambda_{\rm QCD}}}
\def\pr{{\sl Phys. Rev.}~}
\def\prl{{\sl Phys. Rev. Lett.}~}
\def\pl{{\sl Phys. Lett.}~}
\def\np{{\sl Nucl. Phys.}~}
\def\zp{{\sl Z. Phys.}~}
\def\lsim{ {\ \lower-1.2pt\vbox{\hbox{\rlap{$<$}\lower5pt\vbox{\hbox{$\sim$}
}}}\ } }
\def\gsim{ {\ \lower-1.2pt\vbox{\hbox{\rlap{$>$}\lower5pt\vbox{\hbox{$\sim$}
}}}\ } }

\font\el=cmbx10 scaled \magstep2{\obeylines\hfill May, 2010}

\vskip 1.5 cm

\centerline{\large\bf Long-Distance Contributions to $D^0$-$\bar D^0$ Mixing Parameters}
\bigskip
\bigskip
\centerline{\bf Hai-Yang Cheng,$^{1,2}$ Cheng-Wei Chiang$^{3,1}$}
\medskip
\centerline{$^1$ Institute of Physics, Academia Sinica}
\centerline{Taipei, Taiwan 115, Republic of China}
\medskip
\centerline{$^2$ Physics Department, Brookhaven National Laboratory} \centerline{Upton, New York 11973}
\medskip
\centerline{$^3$ Department of Physics and Center for Mathematics } \centerline{and Theoretical Physics, National Central University}
\centerline{Chungli, Taiwan 320, Republic of China}
\medskip

\bigskip
\bigskip
\centerline{\bf Abstract}
\bigskip
\small
Long-distance contributions to the $D^0$-$\bar D^0$ mixing parameters $x$ and $y$ are evaluated using latest data on hadronic $D^0$ decays.  In particular, we take on two-body $D \to PP$ and $VP$ decays to evaluate the contributions of two-body intermediate states because they account for $\sim 50\%$ of hadronic $D^0$ decays.  Use of the diagrammatic approach has been made to estimate yet-observed decay modes.  We find that $y$ is of order a few $\times 10^{-3}$ and $x$ of order  $10^{-3}$ from hadronic $PP$ and $VP$ modes.  These are in good agreement with the latest direct measurement of $D^0$-$\bar D^0$ mixing parameters using the $D^0 \to K_S \pi^+\pi^-$ and $K_S K^+ K^-$ decays by BaBar.  We estimate the contribution to $y$ from the $VV$ modes using the factorization model and comment on the single-particle resonance effects and contributions from other two-body modes involving even-parity states.

\pagebreak

\section{Introduction}

It is well-known that short-distance contributions to the $D^0$-$\bar D^0$ mixing parameters $x$ and $y$ to be defined below are very small, of order $10^{-6}$, owing to the Glashow-Iliopoulos-Maiani (GIM) suppression and the doubly Cabibbo suppression \cite{Cheng:1982,Datta}.\footnote{For a tabulation of predictions on the mixing parameters $x$ and $y$ within and beyond the standard model, see \cite{Petrov:03,Nelson}.}  Since the mixing effects and $CP$ violation in the neutral charmed meson system are expected to be very small compared to kaon and $B$ mesons, it is difficult to observe them experimentally.  Nevertheless, BaBar \cite{BaBar:DD}, Belle \cite{Belle:DD} and CDF \cite{CDF:DD} have provided compelling evidence for $D^0$-$\bar D^0$ mixing in the past few years.  The current world averages for the {\it CP} allowed case are \cite{HFAG}
\be
x=(0.98^{+0.24}_{-0.26})\% ~, \qquad y=(0.83\pm0.16)\% ~. \qquad 
\en
%
However, the analyses that have reported evidence for mixing have not been able to provide direct measurements of $x$ and $y$.  A most recent BaBar experiment using the $D^0\to K_S^0\pi^+\pi^-$ and $D^0\to K_S^0K^+K^-$ decays measured the mixing parameters $x$ and $y$ directly, with the results \cite{BaBar:DD2010}
\be
x=(1.6\pm2.3\pm1.2\pm0.8)\times 10^{-3} ~, \qquad
y=(5.7\pm2.0\pm1.3\pm0.7)\times 10^{-3} ~,
\en
where the last error comes from the amplitude models used in the analysis.  These results are consistent with the previous Belle measurements from the $K_S^0\pi^+\pi^-$ mode alone \cite{Belle:DD2007}:
\be
x=(8.0\pm2.9^{+0.9+1.0}_{-0.7-1.4})\times 10^{-3} ~, \qquad
y=(3.3\pm2.4^{+0.8+0.6}_{-1.2-0.8})\times 10^{-3} ~.
\en
Therefore, the new BaBar measurement favors lower values for $x$ than for $y$ and lower $x$ and $y$ values than the aforementioned world averages.  At any rate, the observed $D^0$-$\bar D^0$ mixing is several orders of magnitude larger than what is expected from the short-distance contributions, and is evidently dominated by long-distance processes.

Since the long-distance effects are non-perturbative in nature, it is conceivable that it will be very difficult to have a reliable estimate of the charm mixing parameters, especially in view of the fact that the charm quark is not heavy enough for a sensible heavy quark expansion and not light enough for the application of chiral perturbation theory.  In general, the long-distance contributions are estimated in either of the two approaches: inclusive and exclusive.  The ``inclusive" approach relies on the heavy quark expansion dictated by the parameter $1/m_c$ (for a recent study, see \cite{Bobrowski:2010xg}).  In the so-called ``exclusive" approach, on the other hand, one sums over contributions from intermediate hadronic states.  Since there are cancellations among states within a given flavor SU(3) multiplet, as first noticed in \cite{Donoghue:1985hh,Wolfenstein}, one needs to know the contribution of each state with high precision in order to have a trustworthy estimate \cite{Falk:y}.  It can be shown that the mixing parameters $x$ and $y$ vanish in the SU(3) limit.  In the exclusive scenario, this means that the cancellation of the Cabibbo-favored (CF) and doubly-Cabibbo-suppressed (DCS) decays with the contributions from singly-Cabibbo-suppressed (SCS) transitions is perfect in the limit of SU(3) symmetry.  In other words, nonzero values of $x$ and $y$ comes from the breakdown of flavor SU(3) symmetry.  In the absence of sufficiently precise data on many decays rates and on strong phases, the authors of \cite{Falk:y} computed the long-distance contributions to $\Delta\Gamma$ by considering SU(3) breaking from final-state phase space differences and neglecting SU(3) violation in the decay amplitudes.  They found that the phase space effects alone could produce sufficient SU(3) breaking to induce $y\sim 10^{-2}$ and that large effects in $y$ appeared in decays to final states close to the $D$ threshold.

In the past few years, rich data on hadronic $D\to PP,VP$ decays have been accumulated from various experiments with improved accuracy.  (Throughout the paper, we use $P$, $V$, $S$, $A$, and $T$ to denote pseudoscalar, vector, scalar, axial-vector, and tensor mesons, respectively.)  Especially, there are new CLEO measurements for the $PP$ modes with better precision, many of them having experimental errors less than the present world averages \cite{CLEOPP09}.  These data allow us to make a sensible determination of the mixing parameters $x$ and $y$ from the $PP$ and $VP$ channels without relying on model assumptions. \footnote{For early attempts to estimate the long-distance contributions to $x$ or $y$ from two-body states, see \cite{Colangelo,Kaeding,Buccella96}.}  There are some channels that have not been measured: six SCS $VP$ modes and many of DCS $PP$ and $VP$ decays.  We will employ the diagrammatic approach to give inputs for those unmeasured channels.

Recently, we have studied the two-body hadronic charmed meson decays, including all the $PP$, $VP$, $SP$, $AP$ and $TP$ modes, within the diagrammatic and factorization approaches \cite{CC:charm,CC:pwave}.  The best-fitted values extracted from the CF decay modes in the diagrammatic approach have been used to predict the branching fractions of the SCS and DCS modes for the $D\to PP$ and $D\to VP$ decays.  This approach enables us to estimate the mixing parameters $x$ and $y$.

The layout of this work is as follows.  In Section~II, we first review the method of computing $y$ from hadronic decay branching fractions, and then consider various two-body intermediate-state contributions to $y$.  In Section~III, we review the dispersion relation for $x$, and compute $PP$ and $VP$ mode contributions to $x$, followed by a brief discussion on single particle effects. We summarize our findings in Section~IV.  Our Appendix~A compares different state normalization conventions commonly seen in the literature.  Appendix~B discusses the relative strong phase between the $D^0 \to K^+ \pi^-$ and $K^+ \pi^-$ modes.

\section{Decay width difference}

The neutral $D$ meson mass eigenstates are defined in terms of flavor eigenstates as
\be
|D_{1,2}\ra=p|D^0\ra\pm q|\ov D^0\ra ~.
\en
Using the mass and width matrices, the ratio $q/p$ reads
\be
{q\over p}=\sqrt{{ M_{12}^*-{i\over 2}\Gamma_{12}^*\over M_{12}-{i\over 2}\Gamma_{12} }} ~,
\en
where the convention $CP|D^0\ra=|\ov D^0\ra$ has been made.  The parameters $x$ and $y$ are defined as
\be
x\equiv {\Delta m\over \Gamma}={m_1-m_2\over\Gamma} ~,  \qquad
y\equiv {\Delta \Gamma\over 2\Gamma}={\Gamma_1-\Gamma_2\over 2\Gamma} ~.
\en

Since $CP$ violation in both $D$ mixing and decays is expected to be small within the standard model and in most new physics scenarios, we therefore define the $CP$ eigenstates
\be
|D_{\pm}\ra = {1\over\sqrt{2}}(|D^0\ra \pm |\overline{D}^0\ra) ~, \quad \mbox{ with } \quad
CP |D_{\pm}\ra = \pm |D_{\pm}\ra ~.
\en
Hence $D_1\approx D_+$ and $D_2\approx D_-$.
In perturbation theory, the off-diagonal mass and width matrix elements are given by \cite{Marshak}\footnote{The expressions of the $D^0$-$\ov D^0$ matrix element given in the literature are often very confusing as they are not dimensionally consistent. This issue is discussed in Appendix A. In practice, we do not use Eq. (\ref{eq:deltam}) to compute the long-distance contributions to $\Delta m$ and $\Delta\Gamma$. Rather, we employ Eq. (\ref{eq:y}) and Eq. (\ref{eq:x}) [or Eq. (\ref{eq:x2body})] to evaluate the parameters $y$ and $x$, respectively. They are free of any ambiguities.}
\be
\left(M-{i\over 2}\Gamma\right)_{12}=
{1\over 2m_D}\la D^0|H_w|\ov D^0\ra+{1\over 2m_D}
\sum_n {1\over \N}{\la D^0|H_w|n\ra\la n|H_w|\ov D^0\ra\over m_D-E_n+i\epsilon} ~,
\en
where $\N$ is an appropriate normalization factor for the intermediate state $|n\ra$; for example, $\N=2E_n$ for a one-particle intermediate state.  Using the relation
\be
{1\over m_D-E_n+i\epsilon}={\cal P}{1\over m_D-E_n}-i\pi\delta(m_D-E_n) ~,
\en
with $\cal P$ denoting the principal value prescription, we have
\be \label{eq:deltam}
\Delta m &=& {1\over m_D}\la D^0|H_w|\ov D^0\ra+{1\over 2m_D}{\cal P}\sum_n {1\over\N}{\la D^0|H_w|n\ra\la n|H_w|\ov D^0\ra+\la \ov D^0|H_w|n\ra\la n|H_w|D^0\ra\over m_D-E_n} ~, \non \\
\Delta \Gamma &=& {1\over 2m_D}\sum_n{1\over\N}\left[ \la D^0|H_w|n\ra\la n|H_w|\ov D^0\ra+\la \ov D^0|H_w|n\ra\la n|H_w|D^0\ra\right]
(2\pi)\delta(m_D-E_n) ~.
\en
Therefore, $\Delta m$ and $\Delta \Gamma$ are induced by off-shell and on-shell intermediate states, respectively.

The parameter $y$ has the expression
\be \label{eq:y}
y &\approx& {\Gamma_+-\Gamma_-\over 2\Gamma}={1 \over 2}\sum_n(\B(D_+\to n)-\B(D_-\to n)) \non \\
 &=&{1\over 2\Gamma}\sum_n\rho_n\left(|\la D_+|H_w|n\ra|^2-|\la D_-|H_w|n\ra|^2\right),
\en
where $\rho_n$ is a phase-space factor.  For example, $\rho_n=p_c/(8\pi m_D^2)$ for the $D\to PP$ decays, with $p_c$ being the center-of-mass momentum of either meson in the final state.  Defining $CP|f\ra=\eta_{\rm CP}|{\bar f}\ra$, $y$ can be recast to \cite{Falk:y}
\be \label{eq:y}
 y&=& {1\over 2\Gamma}\sum_n\rho_n\eta_{\rm CP}(n)(\la D^0|H_w|n\ra\la\bar n|H_w|D^0\ra+ \la D^0|H_w|\bar n\ra\la n|H_w|D^0\ra) \non \\
 &=& \sum_n \eta_{\rm CKM}(n)\eta_{\rm CP}(n)\cos\delta_n\sqrt{\B(D^0\to n)\B(D^0\to\bar n)} ~,
\en
where $\delta_n$ is the strong phase difference between the $D^0\to n$ and $\bar D^0\to n$ amplitudes and $\eta_{\rm CKM}=(-1)^{n_s}$ with $n_s$ being the number of $s$ and $\bar s$ quarks in the final state.  The factor $\eta_{\rm CP}=\pm1$ is well-defined because $|f\ra$ and $|\bar f\ra$ are in the same $SU(3)$ multiplet. This factor is the same for the entire multiplet.

\subsection{$PP$}

Since $CP|\pi^0\ra=-|\pi^0\ra$ and likewise for $\eta,\eta'$, we will choose the convention such that $CP|K^+\ra=-|K^-\ra$ and $CP|K^0\ra=-|\bar K^0\ra$. Because under the $CP$ transformation
\be \label{eq:CP}
CP|M_1M_2\ra=\eta_{\rm CP}(M_1)\eta_{\rm CP}(M_2)(-1)^L|M_1M_2\ra=\eta_{\rm CP}(M_1M_2)|M_1M_2\ra ~,
\en
it is clear that $\eta_{\rm CP}(PP)=1$ for decays into two pseudoscalar mesons.  The parameter $y$ arising from the $PP$ states is
\be \label{eq:yPP}
y_{PP} &=&
\B(\pi^+\pi^-)+\B(\pi^0\pi^0)+\B(\pi^0\eta)+\B(\pi^0\eta')
+\B(\eta\eta)+\B(\eta\eta')+\B(K^+K^-)+\B(K^0\bar K^0) \non \\
&& -2\cos\delta_{K^+\pi^-}\sqrt{\B(K^-\pi^+)\B(K^+\pi^-)}
-2\cos\delta_{ K^0\pi^0}\sqrt{\B(\bar K^0\pi^0)\B(K^0\pi^0)} \non \\ && -2\cos\delta_{K^0\eta}\sqrt{\B(\bar K^0\eta)\B(K^0\eta)}
-2\cos\delta_{K^0\eta'}\sqrt{\B(\bar K^0\eta')\B(K^0\eta')} ~.
\en

To see that $y_{PP}$ vanishes in the SU(3) limit, we should work on the SU(3) singlet state $\eta_0$ and octet states $\pi, K,\eta_8$.  The octet states have the same masses when SU(3) symmetry is exact.  We first write down the general quark-graph amplitudes (see \cite{CC:charm} for details):
\be \label{eq:QDAmp}
 A(D^0\to K^-\pi^+) &=& V_{cs}^*V_{ud}(T+E) ~, \qquad~~~
 A(D^0\to \ov K^0\pi^0) = {1\over\sqrt{2}}V_{cs}^*V_{ud}(C-E) ~, \non \\
 A(D^0\to \ov K^0\eta_8) &=& {1\over\sqrt{6}}V_{cs}^*V_{ud}(C-E) ~, \quad~
 A(D^0\to \ov K^0\eta_0)={1\over\sqrt{3}}V_{cs}^*V_{ud}(C+2E) ~, \non \\
 A(D^0\to\pi^+\pi^-) &=& V_{cd}^*V_{ud}(T'+E') ~, ~\qquad
 A(D^0\to\pi^0\pi^0) = {1\over \sqrt{2}}V_{cd}^*V_{ud}(C'-E') ~, \non \\
 A(D^0\to\pi^0\eta_8) &=& -{1\over \sqrt{3}}V_{cd}^*V_{ud}E'
   -{1\over \sqrt{3}}V_{cs}^*V_{us}C' ~,\non \\
 A(D^0\to\pi^0\eta_0) &=& -{2\over \sqrt{6}}V_{cd}^*V_{ud}E'
   +{1\over \sqrt{6}}V_{cs}^*V_{us}C' ~, \non \\
 A(D^0\to\eta_8\eta_8) &=& {\sqrt{2}\over 6}V_{cd}^*V_{ud}(C'+E')
   +{\sqrt{2}\over 6}V_{cs}^*V_{us}(-2C'+4E') ~, \non \\
 A(D^0\to\eta_8\eta_0) &=& -{\sqrt{2}\over 3}V_{cd}^*V_{ud}(C'+E')
   +{1\over 3\sqrt{2}}V_{cs}^*V_{us}(C'+4E') ~, \non \\
 A(D^0\to K^+ K^-) &=& V_{cs}^*V_{us}(T'+E') ~, \qquad~~~
 A(D^0\to K^0\ov K^0) = V_{cd}^*V_{ud}E'_s+V_{cs}^*V_{us}E'_d ~, \non \\
 A(D^0\to K^+\pi^-) &=& V_{cd}^*V_{us}(T''+E'') ~, \qquad~~
 A(D^0\to  K^0\pi^0) = {1\over\sqrt{2}}V_{cd}^*V_{us}(C''-E'') ~, \non \\
 A(D^0\to K^0\eta_8) &=& {1\over\sqrt{6}}V_{cd}^*V_{us}(C''-E'') ~, \quad
 A(D^0\to K^0\eta_0)={1\over\sqrt{3}}V_{cd}^*V_{us}(C''+2E'') ~,
\en
where $T,C,E$ are color-allowed, color-suppressed and $W$-exchange amplitudes, respectively.  We have followed the conventional practice to denote the primed amplitudes for the SCS modes and double-primed amplitudes for the DCS decays.  In the SU(3) limit, the primed and unprimed amplitudes should be the same, the strong phase $\delta_n$ vanishes, and the $D^0\to K^0\ov K^0$ decay is prohibited.  It is easily seen that perfect cancellation occurs among the SU(3) octet final states, as well as among the decay modes $\pi^0\eta_0$, $\eta_8\eta_0$, $\ov K^0\eta_0$ and $K^0\eta_0$ involving the SU(3) singlet $\eta_0$.  Therefore, $y_{PP}$ indeed vanishes in the SU(3) limit.

Flavor SU(3) symmetry breaking occurs in both the decay matrix elements and in the final-state phase space.  Phase space is the only source of SU(3) violation considered in the previous analysis of \cite{Falk:y}, given the paucity of data at that time.  Since all the data of $D\to PP$ are now available except for three of the DCS modes, we can have a much more accurate estimate of $y_{PP}$ directly from the data.  For the yet to be measured DCS modes, we can rely on the relations, for example, $\Gamma(D^0\to K^0\pi^0)/\Gamma(D^0\to \ov K^0\pi^0)=\tan^4\theta_C$ that has been tested in the measurement of the quantity
\be
R(D^0) \equiv {\Gamma(D^0\to K_S\pi^0)-\Gamma(D^0\to K_L\pi^0)
\over \Gamma(D^0\to K_S\pi^0)+\Gamma(D^0\to K_L\pi^0)} ~.
\en
The prediction $R(D^0)=2\tan^2\theta_C=0.107$ agrees quite well with the experimental value of $0.108\pm0.025\pm0.024$ by CLEO \cite{CLEO:RD0}.

From the model-independent analysis in the diagrammatic approach, it has been observed that sizable violation of flavor SU(3) symmetry occurs in some of SCS modes.  The most noticeable example is the ratio $R=\Gamma(D^0\to K^+K^-) / \Gamma(D^0\to\pi^+\pi^-) \approx 2.8$.  If the SU(3) symmetry breaking manifested only in the phase space, one would obtain $R=0.86$ and hence the $K^+K^-$ production rate should be smaller than the $\pi^+\pi^-$ one.  This is in sharp disagreement with experiment.  We have shown in \cite{CC:charm} that in addition to the SU(3) breaking effect in the spectator amplitudes, the long-distance resonant contribution through the nearby resonance $f_0(1710)$ can naturally explain why $D^0$ decays more copiously to $K^+ K^-$ than $\pi^+ \pi^-$ through the $W$-exchange topology.  This has to do with the dominance of the scalar glueball content of $f_0(1710)$ and the chiral-suppression effect in the decay of a scalar glueball into two pseudoscalar mesons.  The same final-state interaction (FSI) also explains the measured rate of $D^0\to K^0\bar K^0$ even though its amplitude vanishes in the SU(3) limit.

CLEO has measured the relative strong phase between $D^0\to K^+\pi^-$ and $D^0\to K^-\pi^+$ to be $\cos\delta=1.03^{+0.31}_{-0.17}\pm0.06$ \cite{CLEO:phase}.  (See Appendix~B for an estimate of this strong phase.)  It is thus plausible to assume $\cos\delta_n=1$ for all $n=PP$.  From Eq.~(\ref{eq:yPP}) and the data in Table~\ref{tab:PPVP}, we obtain
\be \label{eq:yPPvalue}
y_{PP}=(1.128\pm0.038)\%-(1.042\pm0.017)\%=(0.086\pm 0.041)\% ~.
\en

\begin{table}[t]
\caption{Experimental branching fractions for Cabibbo-favored (in units of \%), singly-Cabibbo-suppressed (in units of $10^{-3}$) and doubly-Cabibbo-suppressed decays (in units of $10^{-3}$) of $D^0\to PP, VP$.  Data are taken from \cite{CLEOPP09} for $D\to PP$ and from \cite{PDG} for $D\to VP$.  The channels with the superscript $*$ have not been measured yet.  For them, we use the fitted branching fractions obtained from the diagrammatic approach \cite{CC:charm}.  For $D\to VP$ decays, the fitted branching fractions are those obtained from solution (A,A1) (upper row) and (S,S1) (lower row) in \cite{CC:charm}. \label{tab:PPVP}
}
\vspace{6pt}
\begin{ruledtabular}
\begin{tabular}{l c | l c | l c}
Mode & ${\cal B}_{\rm exp}$ &  Mode & ${\cal B}_{\rm exp}$ & Mode & ${\cal B}_{\rm exp}$\\
\hline
$K^{-} \pi^+$ & $(3.91 \pm 0.08)\%$ &
$K^{*-} \pi^+$ & $(5.91 \pm 0.39)\%$ &
$\pi^0 \phi$ & $(1.24 \pm 0.12)\times 10^{-3}$
\\
$\ol{K}^{0} \pi^0$ & $(2.38 \pm 0.09)\%$ &
$\ol{K}^{*0} \pi^0$ & $(2.82 \pm 0.35)\%$ &
$\eta \omega$ & $(2.21\pm0.23)\times 10^{-3}$
\\
$\ol{K}^{0} \eta$ & $(0.96 \pm 0.06)\%$ &
$K^- \rho^+$ & $(10.8 \pm 0.7)\%$ &
$\eta\,' \omega$ & $\cases{(0.07 \pm 0.02)\times 10^{-3}$$^* \cr
(0.15 \pm 0.01)\times 10^{-3}$$^*}$
\\
$\ol{K}^{0} \eta\,'$  & $(1.90\pm0.11)\%$ &
$\ol{K}^0 \rho^0$     & $(1.54\pm0.12)\%$ &
$\eta \phi$ & $(0.14\pm0.05)\times 10^{-3}$
\\
$\pi^+ \pi^-$ & $(1.45 \pm 0.05)\times 10^{-3}$ &
$\ol{K}^{*0} \eta$ & $(0.96 \pm 0.30)\% $ &
$\eta \rho^0$ & $\cases{(1.11 \pm 0.86)\times 10^{-3}$$^* \cr
(1.17 \pm 0.34)\times 10^{-3}$$^*}$
\\
$\pi^0 \pi^0$ & $(0.81 \pm 0.05)\times 10^{-3}$ &
$\ol{K}^{*0} \eta\,'$ & $(0.012 \pm 0.003)\%$ &
$\eta\,' \rho^0$ & $\cases{(0.14 \pm 0.02)\times 10^{-3}$$^* \cr
(0.26 \pm 0.02)\times 10^{-3}$$^*}$
\\
$\pi^0 \eta $ & $(0.68 \pm 0.07)\times 10^{-3}$ &
$\ol{K}^0 \omega$ & $(2.26 \pm 0.40)\%$ &
$K^{*+}\,\pi^-$ & $(3.54^{+1.80}_{-1.05})\times 10^{-4}$
\\
$\pi^0 \eta' $ & $(0.91 \pm 0.13)\times 10^{-3}$  &
$\ol{K}^0 \phi$ & $(0.868 \pm 0.060)\%$ &
$K^{*0}\,\pi^0$ & $\cases{(0.54 \pm 0.18)\times 10^{-4}$$^* \cr
(0.74 \pm 0.17)\times 10^{-4}$$^*}$
\\
$\eta\eta $ & $(1.67 \pm 0.18)\times 10^{-3}$ &
$\pi^+ \rho^-$ & $(4.97 \pm 0.23)\times 10^{-3}$ &
$\rho^-\,K^+$ & $\cases{(1.45 \pm 0.17)\times 10^{-4}$$^* \cr
(1.91 \pm 0.21)\times 10^{-4}$$^*}$
\\
$\eta\eta' $ & $(1.05 \pm 0.26)\times 10^{-3}$  &
$\pi^- \rho^+$ & $(9.8 \pm 0.4)\times 10^{-3}$ &
$\rho^0\,K^0$ & $\cases{(0.91 \pm 0.51)\times 10^{-4}$$^* \cr
(0.63 \pm 0.19)\times 10^{-4}$$^*}$
\\
$K^+ K^{-}$ & $(4.07 \pm 0.10)\times 10^{-3}$  &
$\pi^0 \rho^0$ & $(3.73 \pm 0.22)\times 10^{-3}$ &
$K^{*0}\,\eta$ & $\cases{(0.33 \pm 0.08)\times 10^{-4}$$^* \cr
(0.28 \pm 0.05)\times 10^{-4}$$^*}$
\\
$K^0 \ol{K}^{0}$ & $(0.64\pm0.08)\times 10^{-3}$  &
$K^+ K^{*-}$ & $(1.53 \pm 0.15)\times 10^{-3}$ &
$K^{*0}\,\eta^{\prime}$ & $\cases{(0.0040 \pm 0.0006)\times 10^{-4}$$^* \cr
(0.0061 \pm 0.0004)\times 10^{-4}$$^*}$
\\
$K^{+} \pi^-$ & $(1.48 \pm 0.07)\times 10^{-4}$  &
$K^- K^{*+}$ & $(4.41 \pm 0.21)\times 10^{-3}$ &
$\omega\,K^0$ & $\cases{(0.58 \pm 0.40)\times 10^{-4}$$^* \cr
(0.85 \pm 0.21)\times 10^{-4}$$^*}$
\\
${K}^{0} \pi^0$ & $(0.67 \pm 0.02)\times 10^{-4}$$^*$ &
$K^0 \ol{K}^{*0}$ & $\cases{(0.05 \pm 0.06)\times 10^{-3}$$^* \cr
(0.29 \pm 0.22)\times 10^{-3}$$^*}$ &
$\phi\,K^0$ & $\cases{(0.06 \pm 0.05)\times 10^{-4}$$^* \cr
(0.15 \pm 0.06)\times 10^{-4}$$^*}$
\\
${K}^{0} \eta$ &  $(0.28 \pm 0.02)\times 10^{-4}$$^*$ &
$\ol{K}^0 K^{*0}$ & $\cases{(0.29 \pm 0.22)\times 10^{-3}$$^* \cr
(0.05 \pm 0.06)\times 10^{-3}$$^*}$
\\
${K}^{0} \eta\,'$ &  $(0.55 \pm 0.03)\times 10^{-4}$$^*$ &
$\pi^0 \omega$ & $\cases{(0.10 \pm 0.18)\times 10^{-3}$$^* \cr
(1.01 \pm 0.18)\times 10^{-3}$$^*}$
\\
\end{tabular}
\end{ruledtabular}
\end{table}

\subsection{$VP$}

The neutral vector mesons $\rho^0,\omega,\phi$ are $CP$ eigenstates with $CP=+$.  It is thus convenient to define $CP|V\ra=|\ov V\ra$ for the vector mesons in the same $SU(3)$ multiplet.  It follows from Eq.~(\ref{eq:CP}) that $\eta_{\rm CP}(VP)=+1$ for decays into one vector meson and one pseudoscalar meson.  There are more decay modes available for the $VP$ final states, namely, $V_1P_2$ and $P_1V_2$.  There are totally 30 $VP$ channels (8 for CF, 14 for SCS and 8 for DCS), to be compared with 16 $PP$ channels.  Because the decay constant of the vector meson $f_V$, typically of order 210 MeV, is much larger than $f_P$, many $VP$ modes have rates greater than the $PP$ ones.  For example, $\B(K^-\rho^+)\sim 11\%\gg \B(K^-\pi^+)\sim 4\%$ and $\B(\pi^-\rho^+)\sim 1\%\gg \B(\pi^+\pi^-)\sim 1.5\times 10^{-3}$.  It is thus anticipated that the $VP$ mode contributions to $y$ will dominate over $y_{PP}$.

Note that the decay amplitude of the DCS mode is not simply related to the corresponding CF one by replacing the CKM matrix elements $V_{cs}^*V_{ud}$ with $V_{cd}^*V_{us}$. For example,
\be
{A(D^0\to K^{*+}\pi^-)\over A(D^0\to K^{*-}\pi^+)}={V_{cd}^*V_{us}\over V_{cs}^*V_{ud}}\,{T_P+E_V\over T_V+E_P} ~,
\en
where the subscripts indicate which final-state meson contains the spectator quark in the $D$ meson.  In the diagrammatic approach, amplitudes of the same topology by different subscripts are {\it a priori} independent of each other.  In the SU(3) symmetry limit, however, they are identical.  It is also instructive to see that in this case, the $VP$ contribution to $y$ in each SU(3) multiplet also vanishes.

In \cite{CC:charm}, we obtain two possible solutions, called (S,S1) and (A,A1), for the $T_{V,P}$, $C_{V,P}$, and $E_{V,P}$ amplitudes, depending on which formula is used to extract the invariant amplitudes.\footnote{The solution (S,S1) is obtained by using the equation
 \be \label{eq:VP1}
 \Gamma(D\to VP)={p_c\over 8\pi m_D^2}\sum_{pol.}|\A|^2 \non
 \en
to extract the invariant amplitude, while the relation
 \be \label{eq:VP2}
 \Gamma(D\to VP)={p^3_c\over 8\pi m_D^2}|\tilde \A|^2, \non
 \en
is used to get the solution (A,A1), where the polarization vector is taken out of the amplitude so that $\A= (m_V/m_D)\tilde \A\, (\varepsilon\cdot p_D)$. \label{ftnt:VP}}
The ones quoted in Table~\ref{tab:PPVP} are from the (A,A1) solution (upper row) and the (S,S1) solution (lower row).  We find
\be \label{eq:yVP}
y_{VP}=\cases{
(2.847\pm0.112)\%-(2.578\pm0.227)\%=(0.269\pm 0.253)\% &
{(A,A1)} \cr
(2.916\pm0.073)\%-(2.764\pm0.207)\%=(0.152\pm 0.220)\% &
{(S,S1)}}
\en
where the error bars are of the same order as the central values.  As far as the central value is concerned, $y_{VP}$ is indeed larger than $y_{PP}$.

\subsection{$VV$}

Just as the $PP$ modes, there are 16 $VV$ channels: 4 for CF: $K^{*-}\rho^+,\ov K^{*0}\rho^0,\ov K^{*0}\omega,\ov K^{*0}\phi$; 8 for SCS: $K^{*+}K^{*-},K^{*0}\ov K^{*0},\rho^+\rho^-,\rho^0\rho^0,\rho^0\omega,\rho^0\phi,\omega\omega,\omega\phi$ and 4 for DSC: $K^{*+}\rho^-, K^{*0}\rho^0,K^{*0}\omega, K^{*0}\phi$.  Among them, $\ov K^{*0}\phi$ and $K^{*0}\phi$ are kinematically forbidden, but allowed through the finite width effect of $K^*$.  Indeed, the decay $D^0\to \ov K^{*0}\phi$ has been observed by FOCUS in the Dalitz plot analysis of $D^0\to K^+K^-K^-\pi^+$ \cite{phiKst}.

The measurements of $K^{*-}\rho^+,\ov K^{*0}\rho^0,\ov K^{*0}\omega$ were performed in the early 90's.  During the period of 2003-2007, FOCUS had measured $\bar K^{*0}\phi, \rho^0\phi,K^{*0}\ov K^{*0}$ and $\rho^0\rho^0$ through the Dalitz plot analysis of various four-body decay modes \cite{phiKst,FOCUS}.  Some of the $VV$ data are problematic.  Na{\"i}vely, it is expected from the factorization hypothesis that longitudinal and transverse polarizations are comparable in $D\to VV$.  However, the Mark~III measurement \cite{MarkIII} has indicated that the branching fraction of the $D^0\to \bar K^{*0}\rho^0$ decay is already saturated by the transverse polarization state (see, {\it e.g.}, \cite{Kamal99} for a detailed discussion).  Since the presently available data do not allow us to have a sensible determination of $y_{VV}$, we will rely on the factorization model to estimate its magnitude.

Note that the polarized decay amplitudes can be expressed in several different but equivalent bases.  For example, the helicity amplitudes can be related to the spin amplitudes in the transversity basis $(A_0, A_\|, A_\bot)$ defined in terms of the linear polarization of the vector mesons, or to the partial-wave amplitudes $(S,P,D)$ via:
 \be
A_0 &=& H_{0}= -{1\over\sqrt{3}}\, S+\sqrt{2\over 3}\, D ~, \non \\
A_\| &=& {1\over\sqrt{2}}(H_{+}+ H_{-})=\sqrt{2\over 3}
\,S+{1\over\sqrt{3}}\,D ~, \non \\
A_\bot &=& {1\over\sqrt{2}}(H_{+}- H_{-})=P ~,
 \en
where we have followed the sign convention of \cite{Dighe}.  The decay rate reads
\be
\Gamma(D\to V_1V_2) &=& {p_c\over
8\pi m_D^2}(|H_{0}|^2+|H_{+}|^2+|H_{-}|^2) ~, \non \\
&=& {p_c\over 8\pi m_D^2}(|A_{0}|^2+|A_\bot|^2+|A_\||^2) ~, \non \\
&=& {p_c\over 8\pi
m_D^2}(|S|^2+|P|^2+|D|^2) ~.
\en
\begin{table}[t]
\caption{Branching fractions of $D^0\to VV$ calculated in the factorization approach.  Data are taken from \cite{CLEOPP09}.  Since the $W$-exchange contributions are neglected in na{\"i}ve factorization, no estimate of the branching fractions of $K^{*0}\ov K^{*0},\ov K^{*0}\phi,K^{*0}\phi$ and $\rho^0\omega$ is made here.
\label{tab:DVV}}
\vspace{6pt}
\begin{ruledtabular}
\begin{tabular}{l c c c c c c }
Mode & $S$ wave &  $P$ wave & $D$ wave & $\B_{\rm total}$ & Expt.  \\
\hline
$K^{*-} \rho^+$ & 10.5\% & $6.5\times 10^{-3}$ & $1.6\times 10^{-3}$ & $11.3\%$& $(6.5\pm2.5)\%$ \\
$\ov K^{*0} \rho^0$ & 1.64\% & $1.4\times 10^{-3}$ & $2.3\times 10^{-4}$ & 1.8\% & $(1.59\pm0.35)\%$ \\
$\ov K^{*0} \omega$ & 1.5\% & $1.2\times 10^{-3}$ & $1.8\times 10^{-4}$ & 1.6\% & $(1.1\pm0.5)\%$ \\
$K^{*+} K^{*-}$ & $6.8\times 10^{-3}$ & $5.5\times 10^{-4}$ & $8.9\times 10^{-8}$ & $7.3\times 10^{-3}$ & $$ \\
$\rho^+\rho^-$ & $5.8\times 10^{-3}$ & $6.2\times 10^{-4}$ & $2.3\times 10^{-4}$ & $6.6\times 10^{-3}$ & $$ \\
$\rho^0\rho^0$ & $0.85\times 10^{-3}$ & $0.91\times 10^{-4}$ & $3.4\times 10^{-5}$ & $0.97\times 10^{-3}$ & $(1.83\pm0.13)\times 10^{-3}$ \\
$\rho^0\phi$ & $6.3\times 10^{-4}$ & $2.5\times 10^{-5}$ & $1.0\times 10^{-6}$ & $6.6\times 10^{-4}$ \\
$\omega\omega$ & $5.9\times 10^{-4}$ & $6.5\times 10^{-5}$ & $1.9\times 10^{-5}$ & $6.8\times 10^{-4}$ \\
$\omega\phi$ & $6.3\times 10^{-4}$ & $2.5\times 10^{-5}$ & $1.0\times 10^{-6}$ & $6.6\times 10^{-4}$ \\
$K^{*+} \rho^-$ & $3.2\times 10^{-4}$ & $2.8\times 10^{-5}$ & $4.5\times 10^{-6}$ & $3.5\times 10^{-4}$& $$ \\
$K^{*0} \rho^0$ & $4.4\times 10^{-5}$ & $2.7\times 10^{-6}$ & $6.7\times 10^{-7}$ & $4.7\times 10^{-5}$ & $$ \\
$K^{*0} \omega$ & $3.3\times 10^{-5}$ & $2.0\times 10^{-6}$ & $4.3\times 10^{-7}$ & $3.5\times 10^{-5}$ & $$ \\
\end{tabular}
\end{ruledtabular}
\end{table}

The factorizable matrix element for the $D\to V_1V_2$ decay is
\be
X_h^{(DV_1,V_2)} &\equiv & \la V_2 |J^{\mu}|0\ra\la
V_1|J'_{\mu}|D \ra =- if_{V_2}m_2\Bigg[
(\vp^*_1\cdot\vp^*_2) (m_{D}+m_{V_1})A_1^{ DV_1}(m_{V_2}^2)  \non \\
&-& (\vp^*_1\cdot p_{_{D}})(\vp^*_2 \cdot p_{_{D}}){2A_2^{
DV_1}(m_{V_2}^2)\over (m_{D}+m_{V_1}) } +
i\epsilon_{\mu\nu\alpha\beta}\vp^{*\mu}_2\vp^{*\nu}_1p^\alpha_{_{D}}
p^\beta_1\,{2V^{ DV_1}(m_{V_2}^2)\over (m_{D}+m_{V_1}) }\Bigg] ~,
\en
where use of the conventional definition for form factors \cite{BSW} has been made.  The longitudinal ($h=0$) and transverse ($h=\pm$) components of $X^{( \bar DV_1,V_2)}_h$ are given by
 \be \label{eq:Xh}
 X_0^{(DV_1,V_2)} &=& {if_{V_2}\over 2m_{V_1}}\left[
 (m_D^2-m_{V_1}^2-m_{V_2}^2)(m_D+m_{V_1})A_1^{DV_1}(q^2)-{4m_D^2p_c^2\over
 m_D+m_{V_1}}A_2^{DV_1}(q^2)\right] ~, \non \\
 X_\pm^{(DV_1,V_2)} &=& -if_{V_2}m_Dm_{V_2}\left[
 \left(1+{m_{V_1}\over m_D}\right)A_1^{DV_1}(q^2)\pm{2p_c\over
 m_D+m_{V_1}}V^{DV_1}(q^2)\right] ~.
 \en

In the factorization framework, we find that $|H_0|^2\sim |H_+|^2> |H_-|^2$ and $|S|^2>|P|^2>|D|^2$.  Therefore, the longitudinal polarization $f_L$ defined by
\be
f_L\equiv \frac{\Gamma_L}{\Gamma}
=\frac{| A_0|^2}{|A_0|^2+|A_\parallel|^2+|A_\bot|^2}=\frac{| H_0|^2}{|H_0|^2+|H_+|^2+|H_-|^2}
\label{eq:fL}
\en
is expected to be in the vicinity of 0.5\,.  Indeed, $f_L=0.475\pm0.271$ was found in $D^0\to K^{*-}\rho^+$ \cite{MarkIII:VV}.  This is not the case in tree-dominated charmful or charmless $B \to VV$ decays where the longitudinal polarization dominates, {\it i.e.}, $|H_0|^2> |H_+|^2> |H_-|^2$ and $f_L=1-{\cal O}(m^2_V/ m^2_B)$.  However, for the $D^0\to \ov K^{*0}\rho^0$ decay, it was found by Mark~III \cite{MarkIII:VV} that this mode proceeded through the transverse polarization, with only a tiny room for the longitudinal polarization.  More precisely, $\B(D^0\to\ov K^{*0}\rho^0)_{\rm trasnverse}=(1.6\pm0.6)\%$, while the total rate is $\B(D^0\to\ov K^{*0}\rho^0)_{\rm tot}=(1.59\pm0.35)\%$.  Mark~III also measured the partial wave branching fractions: $(3.1\pm0.6)\%$, $<3\times 10^{-3}$ and $(2.1\pm0.6)\%$, respectively, for the $S$-, $P$- and $D$-waves \cite{MarkIII:VV,PDG}. The sum of $S$- and $D$-wave branching fractions already exceeds the total. Hence, the data associated with this mode are problematic.

The $VV$ states with different partial waves contribute with different {\it CP} parties.  We have $\eta_{\rm CP}(VV)=1$ for $VV$ in a relative $S$ or $D$ wave, and $-1$ in a $P$ wave \cite{Falk:y}.  The parameter $y$ for $VV$ modes has the expression
\be \label{eq:yVV}
y_{VV,\ell} &=&
\B(\rho^+\rho^-)_\ell+\B(\rho^0\rho^0)_\ell+\B(\rho^0\omega)_\ell
+\B(\rho^0\phi)_\ell+\B(\omega\omega)_\ell+\B(\omega\phi)_\ell
+\B(K^{*+}K^{*-})_\ell \non \\
&&+ \B(K^{*0}\bar K^{*0})_\ell
-2\cos\delta_{K^{*+}\rho^-}\sqrt{\B(K^{*-}\rho^+)_\ell\B(K^{*+}\rho^-)_\ell}
-2\cos\delta_{ K^{*0}\rho^0}\sqrt{\B(\bar K^{*0}\rho^0)_\ell\B(K^{*0}\rho^0)_\ell} \non \\
&&- 2\cos\delta_{K^{*0}\omega}\sqrt{\B(\bar K^{*0}\omega)_\ell\B(K^{*0}\omega)_\ell}
-2\cos\delta_{ K^{*0}\phi}\sqrt{\B(\bar K^{*0}\phi)_\ell\B(K^{*0}\phi)_\ell}
\en
for $\ell=S,D$, and an overall minus sign is needed for $\ell=P$.

Using the effective Wilson coefficients $a_1=1.22$ and $a_2=-0.66$ \cite{CC:charm}, the form factors from the covariant light-front quark model \cite{CCH} and the decay constants \cite{BallfV}
\be
f_\rho=216\,{\rm MeV} ~, \quad f_{K^*}=220\,{\rm MeV} ~, \quad
f_{\phi}=215\,{\rm MeV} ~, \quad f_{\omega}=187\,{\rm MeV} ~,
\en
the predicted branching fractions of $D\to VV$ decays for various partial waves within the factorization framework are shown in Table~\ref{tab:DVV}.  Since the $W$-exchange contributions are neglected in na{\"i}ve factorization, no estimate of the branching fractions of $K^{*0}\ov K^{*0},\ov K^{*0}\phi,K^{*0}\phi$ and $\rho^0\omega$ is made here.  We obtain
\be
y_{_{VV}}^{S-wave}=0.06\% ~,\qquad y_{_{VV}}^{P-wave}=-0.03\% ~, \qquad y_{_{VV}}^{D-wave}=0.007\% ~.
\en
A cancellation between even-parity and odd-parity final states renders small $y_{VV}$ in comparison with $y_{PP,VP}$.

\subsection{$D\to MP$}

Recently we have studied the hadronic $D$ meson decays into a pseudoscalar meson $P$ and an even-parity meson $M$, where $M$ represents a scalar meson $S$, an axial-vector meson $A$, or a tensor meson $T$ \cite{CC:pwave}.  The data are inferred from detailed Dalitz plot analyses of three-body or four-body decays.  Normally one applies the narrow width approximation
\be \label{eq:fact}
 \Gamma(D\to MP\to P_1P_2P)=\Gamma(D\to MP)\B(M\to P_1P_2)
\en
to extract the branching fractions of the $D\to MP$ decays.  There are two complications though: (i) Some decays occur near or slightly above the threshold, for example, $D^0\to f_0(980)\pi^0$ followed by $f_0\to K^+K^-$.  Since the central values of the $f_0(980)$ and $a_0(980)$ masses are below the threshold for decaying into a pair of charged kaons, one cannot apply the narrow width approximation to extract $\B(D^0\to f_0(980)\pi^0)$ from $D^0\to f_0(980)\pi^0$ followed by $f_0\to K^+K^-$. (ii) Some states, {\it e.g.}, $\sigma$ or $\kappa$ are very broad in their widths. The use of the narrow width approximation is not justified, and it becomes necessary to take into account the finite width effect of broad resonances.  For example, we find that the branching fraction of $D^+\to\sigma\pi^+$ extracted from three-body decays is enhanced by a factor of 2, whereas $\B(D^0\to f_2(1270)\ov K^0)$ is reduced by a factor of 4 by finite width effects \cite{CC:pwave}.  The current experimental data for the two-body decays of $D^0$ to $SP,AP$ and $TP$ are collected in Tables~\ref{tab:SATPdata}. Evidently, one cannot use these data to predict $y$ at this stage.

\begin{table}[t]
\caption{Experimental branching fractions for Cabibbo-favored (upper portion) and singly-Cabibbo-suppressed (lower portion) decays of $D^0\to SP,AP,TP$ (see \cite{CC:pwave} for details).
\label{tab:SATPdata}}
\vspace{6pt}
\begin{ruledtabular}
\begin{tabular}{l c | l c | l c}
Mode & ${\cal B}_{\rm exp}$ &  Mode & ${\cal B}_{\rm exp}$ & Mode & ${\cal B}_{\rm exp}$\\
\hline
$a_0^0\ol{K}^{0}$
     & $(1.6 \pm 0.5)\%$  & $K_1^-(1270) \pi^+$
     & $(1.14 \pm 0.32)\%$ &  $K_2^{*-}\pi^+$
     & $(2.1^{+1.2}_{-0.7})\times 10^{-3}$  \\
$f_0\ov K^0$
     & $(8.0^{+2.5}_{-2.2})\times 10^{-3}$    & $K^-a_1^+(1260)$
     & $(7.9 \pm 1.1)\%$ &  $f_2\ov K^0$
     & $(5.0^{+3.5}_{-2.1})\times 10^{-4}$  \\
${K}_0^{*-} \pi^+$
     & $(8.2 \pm 1.4)\times 10^{-3}$  & & & \\
$\ol{K}_0^{*0} \pi^0$ & $(9.2 ^{+8.1}_{-2.6})\times 10^{-3}$  & & & \\ \hline
$f_0 \pi^0$
     & $(1.0 \pm 0.3)\times 10^{-4}$ & $\pi^-a_1^+(1260)$
     & $(8.98 \pm 0.62)\times 10^{-3}$ & $f_2\pi^0$ &  $(3.4 \pm 0.4)\times 10^{-4}$ \\
$\sigma \pi^0$
     & $(1.8 \pm 0.3)\times 10^{-4}$  & $K_1^\pm(1270)K^\mp$
     & $(8.1\pm1.8)\times 10^{-4} $  &   \\
\end{tabular}
\end{ruledtabular}
\end{table}

The flavor diagram approach and the factorization calculation have been undertaken to analyze these decay processes in \cite{CC:pwave}.  While factorization works well for the CF $D^+ \to SP, AP$ decays, predictions are typically about one order of magnitude smaller than experiment for the other decay modes, conceivably due to the negligence of weak annihilation contributions arising from FSIs.  The $D \to TP$ measurements poise the biggest problem for theory.  Predicted branching fractions based on factorization are at least two orders of magnitude smaller than data, even when the decays are free of weak annihilation contributions.  We cannot find possible sources of rate enhancement for $D\to TP$.

There are $D\to MP$ decays which are not kinematically allowed but may proceed through the final widths of even-parity resonances.  Examples are $D^0\to a_0^\pm(1450)K^\mp$ and $D^0\to K_1^\pm(1400)K^\mp$.  They are needed to ensure the cancellation for $y$ in the SU(3) limit.  Beyond the SU(3) symmetry, these channels are slightly above the threshold and proceed via finite widths.  In this case, SU(3) cancellation may be less effective.

\subsection{Remarks}

Thus far we have concentrated on physical two-body intermediate states.  In principle we should also consider many body final states.  However, we notice that many of 3-body final states arise from $SP,VP,TP$ decays, and 4-body states from $VV,AP$ decays.  Empirically, we know that non-resonant 3-body and 4-body decays are at most 10\% of the multi-body decay rate.

Summing over the currently available data listed in Tables~\ref{tab:PPVP}-\ref{tab:SATPdata}, we have $\B(D^0\to PP)\sim 10\%$, $\B(D^0\to VP)\sim 28\%$, $\B(D^0\to VV)\sim 10\%$, $\B(D^0\to SP)\sim 4.2\%$, $\B(D^0\to AP)\sim 10\%$, $\B(D^0\to TP)\sim 0.3\%$.  Hence, the total branching fraction of two-body hadronic decays is $63\%$.  This is about $3/4$ of the total hadronic rate, recalling that the semileptonic decays account for $16\%$ of the total rate \cite{PDG}.  This means that, unlike the case of $B$ mesons, the hadronic charm decays are dominated by exclusive two-body processes.  Since $PP$ and $VP$ final states account for nearly half of the hadronic width of $D^0$ and $y_{PP+VP}= (0.36\pm0.26)\%$ or $(0.24\pm0.22)\%$, it is conceivable that when all hadronic states are summed over, one could have $y\sim (0.5-0.7)\%$.

\section{Mass difference}

\begin{figure}[t]
\begin{center}
\includegraphics[width=0.4\textwidth]{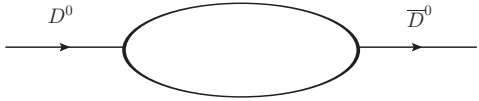}
\vspace{0.0cm}
\caption{Two-particle contribution to the neutral charmed meson mass difference. } \label{fig:selfenergy} \end{center}
\end{figure}

Since the short-distance contribution to $x$ is very small \cite{Cheng:1982,Datta,Golowich:2005}, it is natural to turn to the long-distance effects given by Eq.~(\ref{eq:deltam}):
\be
\Delta m={1\over 2m_D}\,{\cal P}\sum_n{1\over \N}{\la D^0|H_w|n\ra\la n|H_w|\ov D^0\ra+ \la \ov D^0|H_w|n\ra\la n|H_w|D^0\ra \over m_D-E_n} ~.
\en
In principle, one can have one-, two-, three-,... particle intermediate states.  Unlike the width difference, the intermediate states here can be off-shell.  The two-body hadronic intermediate state contributions to $\Delta m$ were first considered in \cite{Wolfenstein,Donoghue:1985hh} by computing the self-energy diagram depicted in Fig.~\ref{fig:selfenergy}.  As stressed in \cite{Donoghue:1985hh}, the self-energy diagram has a universal imaginary part in the massless limit.  A dispersive relation between $x$ and $y$ has been derived in \cite{Falk:x} in the heavy quark limit
\be
\Delta m=-{1\over 2\pi}{\cal P}\int^\infty_{2m_\pi}dE\left[ {\Delta \Gamma(E)\over E-m_D}+{\cal O}\left({\Lambda_{\rm QCD}\over E}\right)\right] ~.
\en
Neglecting {\it CP} violation in the decay amplitude, a model independent relation \footnote{The exact expression of $\tan\phi$ was first obtained in \cite{Ciuchini}. The approximate formula Eq. (\ref{eq:tanphi}) is recovered in the limit of $1-|q/p|\ll 1$.}
\be \label{eq:tanphi}
{y\over x}={1-|q/p|\over \tan\phi}
\en
was obtained in \cite{Ciuchini,Grossman,Kagan}.

It is convenient to introduce the self-energy correlator
\be
\Pi(p^2) = i\int d^4x e^{i(p-p_D)\cdot x}
\la D^0|{\cal T}[H_w(x)H_w(0)]|\ov D^0\ra ~.
\en
An insertion of a complete set of intermediate states and use of the representation for the step function
\be \label{eq:step}
\theta(t)={1\over 2\pi i}\int {e^{i\omega t}\over \omega-i\epsilon}d\omega
\en
will relate the mass and width differences to the self-energy correlator when $p=p_D$:
\be
\Delta m=-{{\rm Re}[\Pi(m_D^2)]\over 2m_D} ~, \qquad
\Delta \Gamma=-{{\rm Im}[\Pi(m_D^2)]\over 2m_D} ~.
\en
The self-energy correlator respects a dispersion relation
\be
\Pi(p^2)={1\over \pi}\int_{s_0}^\infty
{{\rm Im}[\Pi(s)]ds \over s-p^2+i\epsilon} ~,
\en
where $s_0=(m_1+m_2)^2$ with $m_1$ and $m_2$ being the masses in the loop.  Hence,
\be \label{eq:x}
x = {1\over 2m_D\Gamma_D}
{\cal P}\int_{s_0}^\infty {{\rm Im}[\Pi(s)]ds \over s-p^2} ~.
\en
The absorptive part of $\Pi(s)$ amounts to putting the intermediate states in the self-energy diagram on shell.  Hence, it is proportional to the decay rate of $D^0\to n$.  The result is \cite{Burdman}
\be \label{eq:x2body}
x={m_D\over 4\pi}\sum_n \eta_{\rm CKM}(n)\eta_{\rm CP}(n)\cos\delta_n\sqrt{\B(D^0\to n)\B(D^0\to\bar n)}\, {I(m_1,m_2,\Lambda)\over p_c(n)} ~,
\en
with
\be
I(m_1,m_2,\Lambda)=-{\cal P}\int_{s_0}^{\Lambda^2}{\sqrt{1-{s_0\over s}}\over s-m_D^2}ds ~,
\en
where a cutoff $\Lambda$ has been introduced to render the loop integral finite and use has been made of the formula for two-body decay rates
\be
\B(D^0\to n)={p_c\over 8\pi m_D^2 \Gamma}|\la D^0|H_w|n\ra|^2 ~.
\en
A summation over the polarization states of the vector meson is understood for $VP$ final states.  As shown in \cite{Burdman}, the cutoff scale $\Lambda\sim (2.0-2.2)$ GeV is not far from $m_D$.  Note that the result of \cite{Donoghue:1985hh} is recovered in the zero mass limit of intermediate states.

From Eq.~(\ref{eq:x2body}) and Table~\ref{tab:PPVP}, we obtain
\be \label{eq:xPPVP}
x_{PP}=\cases{ (0.028\pm0.003)\% \cr (0.032\pm 0.005)\% \cr (0.039\pm0.008)\% }, \qquad
x_{VP}=\cases{ (0.064\pm0.009)\% & for $\Lambda=2.2$ GeV ~, \cr
(0.073\pm0.021)\% & for $\Lambda=2.1$ GeV ~, \cr
(0.088\pm0.043)\% & for $\Lambda=2.0$ GeV ~. }
\en
Since we have applied the first equation in footnote~\ref{ftnt:VP} to derive $x_{VP}$, only the solution (S,S1) is relevant for the determination of this parameter.  As a result, $x_{PP+VP}$ is of order $10^{-3}$ with the uncertainty depending on the cutoff scale.  As far as the $PP$ and $VP$ modes are concerned, we find that $x_{PP+VP}$ is smaller than $y_{PP+VP}$.  This can be seen by comparing Eq.~(\ref{eq:x2body}) with Eq.~(\ref{eq:y}).  We see that $x$ is suppressed by a factor of $4\pi$, while the factor $m_D I(m_1,m_2,\Lambda)/p_c$ is maximal for $D^0\to \pi\pi$ and of order $2.5$\,.

Just as the $K_L$-$K_S$ mass difference receives contributions from the one-particle intermediate states such as $\pi$, $\eta$ and $\eta'$, \footnote{It is known that the contributions to the $K_L$-$K_S$ mass difference from the octet states $\pi$ and $\eta_8$ cancel exactly as a consequence of the Gell-Mann-Okubo mass relation, as first noticed in the context of SU(3) chiral perturbation theory (ChPT) \cite{Donoghue84}. A generalization to U(3) ChPT to include the $\eta_8-\eta_0$ mixing effects was discussed in \cite{Gerard}. It turns out that the $\eta'$ one-particle intermediate state is one of the main contributions to the $K_L$-$K_S$ mass difference, besides the short-distance contribution and the long-distance contribution represented by the $\pi$ loop \cite{Buras}.}
it will be interesting to see the single particle effects on charm mixing.  A unique feature of the charm system is that an abundant spectrum of resonances is known to exist at energies close to the mass of the charmed meson.  Indeed, the sizable magnitude of the topological $W$-exchange diagram and its large strong phase determined from experiment are suggestive of nearby resonance effects. Considering a nearby resonance state $R$ with mass $m_R$ and width $\Gamma_R$, its contribution to the mass and width differences is
\be
x_R &=& \eta_R{\la D^0|H_w|R\ra\la R|H_w|\bar D^0\ra \over
m_D\Gamma}\,{m_D^2-m_R^2\over (m_D^2-m_R^2)^2+m_D^2\Gamma_R^2} ~, \non \\
y_R &=& \eta_R{\la D^0|H_w|R\ra\la R|H_w|\bar D^0\ra \over
\Gamma}\,{ \Gamma_R\over (m_D^2-m_R^2)^2+m_D^2\Gamma_R^2} ~, \mbox{ and} \\
\frac{x_R}{y_R} &=&
\frac{m_D^2-m_R^2}{m_D \Gamma_R} ~, \non
\en
where $\eta_R$ is the {\it CP} eigenvalue of the resonance $R$.\footnote{Note that our result of $x_R$ differs from that in \cite{Falk:y} by a factor of 2.}
One needs to know the weak couplings of the $D$ meson to the resonance in order to quantify the resonance contributions to the mixing parameters.  Some crude estimates had been made in \cite{Golowich:pole}.

\section{Conclusions}

Motivated by the possibility of observing new physics effects in $D^0-{\bar D}^0$ mixing, a lot of efforts have been put on experimental determination of the $x$ and $y$ parameters in recent years by the BaBar, Belle, and CDF Collaborations.  Now that the experimental precision on both quantities has reached the level of one per mille, it is timely to scrutinize the SM predictions in order to make meaningful inferences.  The short-distance contributions to $x$ and $y$ in the SM have been found to be several orders of magnitude smaller than the observed values.  In contrast, long-distance effects from exchanges of multiple hadronic particles play a more important role.

Since the sum of all two-body hadronic modes that are available currently accounts for about $63\%$ of the $D^0$ decay branching fraction, it is arguable that these channels dominate and provide a good estimate of the mixing parameters.  Other multi-body hadronic decays are empirically less important, particularly when cancellations among them are taken into account.  With more and better-determined two-body hadronic decay branching fractions over the past few years, we are in a better position to take the exclusive approach to evaluate the long-distance effects.  We find that the primary contributions to these parameters come from the $PP$ and $VP$ modes.

To reduce model dependence, we directly take available experimental data and employ the diagrammatic approach to evaluate the yet-observed decay branching fractions.  From the exchanges of $PP$ and $VP$ intermediate states, Eq.~(\ref{eq:yPPvalue}) for $y_{PP}$, Eq.~(\ref{eq:yVP}) for $y_{VP}$ and Eq.~(\ref{eq:xPPVP}) for $x_{PP}$ and $x_{VP}$ are the main results of this paper.  We obtained that $y \sim$ a few $\times 10^{-3}$ and $x \sim 10^{-3}$, with the latter having a mild dependence on a cutoff scale that is assumed to be around 2 GeV.  Here we have assumed that the relative strong phase between each pair of Cabibbo-favored and doubly Cabibbo-suppressed modes is identically zero, thus maximizing the cancellation.  This is partly justified by the determination of the relative phase between $D^0 \to K^+ \pi^-$ and $K^- \pi^+$ decays by CLEO and a theoretical estimate of the phase.  While inclusive analyses generally render $x \gsim y$, our exclusive calculations indicate that $x$ is smaller than $y$, in good agreement with the latest direct measurements of $D^0$-$\bar D^0$ mixing parameters from the Dalitz plot analysis of $D^0 \to K_S\pi^+\pi^-$ and $K_SK^+K^-$ decays by BaBar.

\section*{Acknowledgments}

One of us (H.-Y.~C.) wishes to thank the hospitality of the Physics Department, Brookhaven National Laboratory.  This research was supported in part by the National Science Council of Taiwan, R.~O.~C.\ under Grant Nos.~NSC~97-2112-M-008-002-MY3, NSC~97-2112-M-001-004-MY3 and in part by the NCTS.

\appendix

\section{Normalization of states}

In the literature, the $D^0$-$\ov D^0$ matrix element is usually taken to be \cite{Donoghue:1985hh,Falk:x}
\be \label{eq:Donoghue}
\left(M-{i\over 2}\Gamma\right)_{12} =
{1\over 2m_D}\la D^0|H_w|\ov D^0\ra
+{1\over 2m_D}\sum_n {\la D^0|H_w|n\ra\la n|H_w|\ov D^0\ra\over m_D-E_n+i\epsilon} ~,
\en
or sometimes \cite{Falk:y}
\be
\left(M-{i\over 2}\Gamma\right)_{12}=
\la D^0|H_w|\ov D^0\ra+\sum_n {\la D^0|H_w|n\ra\la n|H_w|\ov D^0\ra\over m_D^2-E_n^2+i\epsilon} ~,
\en
with different normalizations of the $D$ meson state.  It is easily seen that these expressions are dimensionally inconsistent.  The original formula given by Marshak, Riazuddin and Ryan (MRR) \cite{Marshak}
\be \label{eq:Marshak}
\left(M-{i\over 2}\Gamma\right)_{12}=\la D^0|H_w|\ov D^0\ra+\sum_n {\la D^0|H_w|n\ra\la n|H_w|\ov D^0\ra\over m_D-E_n+i\epsilon}
\en
is dimensionally consistent for one-particle intermediate states.  In the MRR convention, the mass dimension of the meson is $[\phi]=-3/2$.  In the usual normalization convention with $[\phi]=-1$, it is tempting to put factors of $1/2m_D$ on Eq.~(\ref{eq:Marshak}) to arrive at Eq.~(\ref{eq:Donoghue}).  However, an appropriate normalization factor for the intermediate state $|n\ra$ ({\it e.g.}, $1/2E_n$ for one-particle intermediate states) is obviously missing in Eq.~(\ref{eq:Donoghue}).

It is instructive to see the derivation of Eq.~(\ref{eq:Marshak}).  In the MRR convention, the self-energy operator is defined by
\be
\Pi(p^2)=i 2m_D\int d^4x e^{i(p-p_D)\cdot x}
\la D^0|{\cal T}[H_w(x)H_w(0)]|\ov D^0\ra ~.
\en
It has a canonical mass dimension of 2.  An insertion of a complete set of one-particle intermediate states
\be \label{eq:intermediate}
\sum_n\int\frac{d^3q}{(2\pi)^3}\frac{m_n}{E_n}|n\ra\la n|
\en
and use of the representation (\ref{eq:step}) for the step function lead to
\be
{\Pi(p_D^2)\over 2m_D}
&=& -\sum_n\int {d^3q}\frac{m_n}{E_n}
\Bigg[
\delta^3({\vec q}_n-\vec{p}_D )
\frac{\la D^0|H_w|n\ra\la n|H_w|{\bar D}^0\ra}{E_D - E_n  + i\epsilon} \non \\
&& \qquad +\delta^3(\vec{p}_D-{\vec q}_n )
\frac{\la {\bar D}^0|H_w|n\ra\la n|H_w|D^0\ra}{E_D - E_n  + i\epsilon}
\Bigg] ~.
\en
In the rest frame of the $D$ meson, ${\vec p}_D = 0$ and $E_D = m_D$.  Therefore,
\be
{\Pi(m_D^2)\over 2m_D}
= -\sum_n
\left[
\frac{\la D^0|H_w|n\ra\la n|H_w|{\bar D}^0\ra}{m_D - E_n + i\epsilon} +
\frac{\la {\bar D}^0|H_w|n\ra\la n|H_w|D^0\ra}{m_D - E_n + i\epsilon}
\right] ~.
\en
Eq.~(\ref{eq:Marshak}) then follows.  Since the insertion of (\ref{eq:intermediate}) is dimensionless, this renders Eq.~(\ref{eq:Marshak}) dimensionally consistent.  For other normalization conventions and for multiparticle intermediate states, an appropriate normalization factor should be included.

\section{Strong phase of $D^0\to K^+\pi^-$ relative to $D^0\to K^-\pi^+$}

In this appendix we give an estimate of the strong phase $\delta_{K^+\pi^-}$ of $D^0\to K^+\pi^-$ relative to $D^0\to K^-\pi^+$ from the experimental measurement of the former.
From Eq.~(\ref{eq:QDAmp}) we learn that the strong phase $\delta_{K^+\pi^-}$ vanishes in the SU(3) limit as the double-primed amplitudes $T''$ and $E''$ should be the same as unprimed amplitudes $T$ and $E$, respectively.  In the limit of SU(3) symmetry, the prediction $\B(D^0\to K^+\pi^-)=(1.12\pm0.05)\times 10^{-4}$ is slightly smaller than the experimental result of $(1.48\pm0.07)\times 10^{-4}$ \cite{PDG}.  This can be understood as the SU(3) breaking effect in the tree amplitude $T''$.  In the factorization approach, the relevant tree amplitudes read
\be
 T &=& {G_F\over
 \sqrt{2}}a_1\,f_\pi(m_D^2-m_K^2)F_0^{DK}(m_\pi^2) ~,
 \non \\
 T'' &=& {G_F\over
 \sqrt{2}}a_1\,f_K(m_D^2-m_\pi^2)F_0^{D\pi}(m_K^2) ~.
\en
Taking the form factors for $D$ to $\pi$ and $K$ transitions from the recent CLEO-c measurements of $D$ meson semileptonic decays to $\pi$ and $K$ mesons \cite{CLEO:FF}, we find $T''/T=1.23$\,.  From CF $D\to PP$ decays we obtain (in units of $10^{-6}$ GeV) \cite{CC:charm}
\be \label{eq:PP1}
T=3.14\pm0.06 ~, \qquad\quad
E=(1.53^{+0.07}_{-0.08})\,e^{i(122\pm2)^\circ} ~.
\en
Combining the above information, we find that the data of $\B(D^0\to K^+\pi^-)$ can be better fitted by having $E''\approx E e^{i10^\circ}$.  This leads to $\delta_{K^+\pi^-}=\arg[(T''+E'')/(T+E)]\approx -7^\circ$ and $\cos\delta_{K^+\pi^-}\approx 0.99$, consistent with the CLEO measurement of $\cos\delta_{K^+\pi^-}=1.03^{+0.31}_{-0.17}\pm0.06$ \cite{CLEO:phase}.

\end{document}